\documentclass[reprint, superscriptaddress,
 amsmath,amssymb,
 aps,
]{revtex4-2}

\usepackage{graphicx}
\usepackage{dcolumn}
\usepackage{bm}
\usepackage{graphicx}                                   
\usepackage{booktabs}                                   
\usepackage{cleveref}                                   
\crefname{figure}{Fig.}{Figs.}
\Crefname{figure}{Figure}{Figures}
\crefname{equation}{Eq.}{Eqs.}
\Crefname{equation}{Equation}{Equations}
\usepackage{amsmath}                                    
\usepackage{siunitx}                                    
\DeclareSIUnit{\bar}{bar}                               
\usepackage{upgreek}

\begin{document}

\title{Surface Acoustic Wave-Mediated Brillouin Temperature Sensor on Thin-Film Lithium Niobate}

\author{Varun Marathan Kovil}
\affiliation{Institute of Photonics, Leibniz University Hannover, Welfengarten 1A 30167, Hannover, Germany}%
\affiliation{Cluster of Excellence PhoenixD, Leibniz University Hannover, Welfengarten 1A, 30167 Hannover, Germany}
 \affiliation{Max Planck Institute for the Science of Light, Staudtstr. 2, 91058 Erlangen, Germany}
 \email{varun.kovil@iop.uni-hannover.de}

\author{Justus Nölkensmeier}%
\affiliation{Max Planck Institute for the Science of Light, Staudtstr. 2, 91058 Erlangen, Germany}
\affiliation{Department of Physics, Friedrich-Alexander-Universität Erlangen- Nürnberg, Staudtstr. 7, 91058 Erlangen, Germany}%

\author{Gladys Jara-Schulz}
\affiliation{Institute of Photonics, Leibniz University Hannover, Welfengarten 1A 30167, Hannover, Germany}%
\affiliation{Cluster of Excellence PhoenixD, Leibniz University Hannover, Welfengarten 1A, 30167 Hannover, Germany}
 \affiliation{Max Planck Institute for the Science of Light, Staudtstr. 2, 91058 Erlangen, Germany}
 
\author{Pratham Kulkarni}
 
\affiliation{Max Planck Institute for the Science of Light, Staudtstr. 2, 91058 Erlangen, Germany}
\affiliation{Department of Physics, Friedrich-Alexander-Universität Erlangen- Nürnberg, Staudtstr. 7, 91058 Erlangen, Germany}%

\author{Silia Babel}
 \affiliation{Paderborn University, Integrated Quantum Optics, Warburger Str. 100, 33098 Paderborn, Germany}%
\affiliation{Paderborn University, Institute for Photonic Quantum Systems (PhoQS), Warburger Str. 100, 33098 Paderborn, Germany}

\author{Christian Golla}
 \affiliation{Paderborn University, Institute for Photonic Quantum Systems (PhoQS), Warburger Str. 100, 33098 Paderborn, Germany}

\author{Rajkumar Jadhav}
 \affiliation{Institute of Photonics, Leibniz University Hannover, Welfengarten 1A 30167, Hannover, Germany}%
 \affiliation{Cluster of Excellence PhoenixD, Leibniz University Hannover, Welfengarten 1A, 30167 Hannover, Germany}
 \affiliation{Max Planck Institute for the Science of Light, Staudtstr. 2, 91058 Erlangen, Germany}
 \affiliation{Department of Physics, Friedrich-Alexander-Universität Erlangen- Nürnberg, Staudtstr. 7, 91058 Erlangen, Germany}%

\author{Christof Eigner}
 \affiliation{Paderborn University, Institute for Photonic Quantum Systems (PhoQS), Warburger Str. 100, 33098 Paderborn, Germany}

\author{Laura Padberg}
\affiliation{Paderborn University, Integrated Quantum Optics, Warburger Str. 100, 33098 Paderborn, Germany}%
\affiliation{Paderborn University, Institute for Photonic Quantum Systems (PhoQS), Warburger Str. 100, 33098 Paderborn, Germany}

\author{Christine Silberhorn}
 \affiliation{Paderborn University, Integrated Quantum Optics, Warburger Str. 100, 33098 Paderborn, Germany}%
\affiliation{Paderborn University, Institute for Photonic Quantum Systems (PhoQS), Warburger Str. 100, 33098 Paderborn, Germany}

\author{Birgit Stiller}
 \affiliation{Institute of Photonics, Leibniz University Hannover, Welfengarten 1A 30167, Hannover, Germany}%
 \affiliation{Cluster of Excellence PhoenixD, Leibniz University Hannover, Welfengarten 1A, 30167 Hannover, Germany}
 \affiliation{Max Planck Institute for the Science of Light, Staudtstr. 2, 91058 Erlangen, Germany}
 \affiliation{Department of Physics, Friedrich-Alexander-Universität Erlangen- Nürnberg, Staudtstr. 7, 91058 Erlangen, Germany}%

\date{\today}

\begin{abstract}
    The emergence of thin-film lithium niobate (TFLN) has paved the way for realizing strong Brillouin interactions in lithium niobate platforms, enabling new opportunities in integrated photonics. However, fully harnessing stimulated Brillouin scattering (SBS) in TFLN for the plethora of Brillouin applications remains an exciting area of research, with significant opportunities for further exploration and technological advancement. Here, we present the first comprehensive experimental and numerical study of the surface acoustic wave (SAW)-mediated Brillouin temperature sensor on a TFLN-on-insulator chip. The proposed technique directly senses the surrounding environment via SAWs, providing an optical-power-efficient platform compatible with existing integrated photonic architectures. Exploiting the strong intermodal Brillouin interactions in x-cut TFLN waveguides, we report a high Brillouin gain of 24.8$ \pm $1.3\, m$^{-1}$W$^{-1}$ and a temperature coefficient of $-0.73$ $\pm$ 0.02\,MHz/$^\circ$C. Considering the anisotropic nature of lithium niobate, we investigate the temperature dependence of the Brillouin frequency for the propagation directions 0$^\circ$, 10$^\circ$, 20$^\circ$, 30$^\circ$ relative to the crystallographic y-axis. 
\end{abstract}

\maketitle

\section{\label{sec:level1}Introduction}
Stimulated Brillouin scattering (SBS) is one of the strongest nonlinear optical processes, arising from the interaction between two optical waves and an acoustic wave \cite{boyd2008,pant2011,wolff2021}. Over the past decades, it has emerged as a promising candidate for a wide range of applications, including optical communication \cite{merklein2022}, optical memory \cite{zhu2007,stiller2024,dong2015,saffer2025}, microwave photonics \cite{mk2022,eggleton2019}, neuromorphic computing \cite{becker2024,slinkov2025}, sensing \cite{bao2011,sanchez2022,Gal2012,Zarifi2019,Zarifi2018}, biomedical applications \cite{palombo2019} and narrow-linewidth lasers \cite{gundavarapu2019,Kabakova2013,chen2024,Chen2026,gyun2017}, benefiting from its inherent properties like narrow linewidth, GHz range resonance frequency, wavelength transparent nature, and bandwidth reconfigurability. In recent years, SBS has been successfully realized on various integrated photonic platforms, including chalcogenide \cite{eggleton2013}, silicon \cite{Li2023}, silicon nitride \cite{gyger2020}, lithium niobate \cite{ye2025,rodrigues2025,Lisa2025,Yu2025,yang2024,Li20261,haerteishigh2026}, and tantalum pentoxide \cite{liu2026}. Among these, lithium niobate is one of the most well-established material platforms for telecommunication and nonlinear optics applications \cite{Zhu2021}, where nonlinear optical processes such as optical parametric amplification \cite{peng2025}, second harmonic generation \cite{yuan2021}, electro-optic modulation \cite{hu2025,kuttner2026}, and four-wave mixing \cite{zheng2021} have been extensively investigated. Recent demonstrations of SBS on TFLN waveguides unlock new Brillouin-based applications and leverage its existing strengths, creating a robust and versatile platform. 

\begin{figure}[htbp]
\centerline{\includegraphics[width=1\columnwidth, height=0.8\columnwidth]{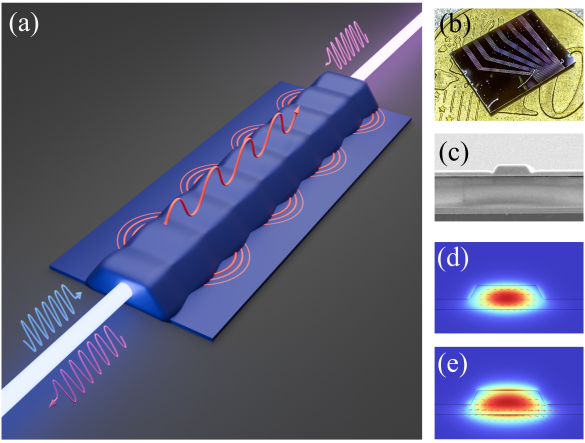}}
\caption{(a) Artistic view of SAWs-mediated Brillouin Stokes scattering in on-chip TFLN waveguide. The counter-propagating pump (blue beam) and probe (purple beam) signals generate a traveling SAWs (red lines and arrows) propagating in the pump direction through electrostriction. The acoustic wave induces a periodic refractive index modulation via the photoelastic effect, enabling Brillouin Stokes scattering; (b) Image of the sample used for the experiments; (c) SEM image of the photonic chip used for experiment; (d) transverse electric (TE) and (e) hybrid modes supported by the waveguide.}
\label{fig:Art}
\end{figure}
\par 
To date, SBS has played a major role in advancing temperature sensing technology by exploiting the linear correlation between the Brillouin resonance frequency and temperature \cite{Gal2012,merklein2022}. Diverse platforms, such as standard single mode fibers (SMF) \cite{Kurashima1990}, photonic crystal fibers (PCF) \cite{Zou2005}, tellurite glass fibers \cite{Mizuno2009}, and polymer optical fibers \cite{OCHI2025} have been explored and studied for the development of stable Brillouin temperature sensor systems with enhanced sensing performance. 
Brillouin temperature sensors can be generally classified into three categories based on the nature of the acoustic waves involved: (1) longitudinal acoustic waves, (2) transverse (or shear) acoustic waves, and (3) surface acoustic waves. SBS mediated by longitudinal acoustic waves is the most common and is typically observed in waveguides exhibiting strong photoelastic coupling \cite{eggleton2013,Huy2016,wolff2021,merklein2022,Kobyakov2010,Ren2025_1}. Since the acoustic waves are strongly confined within the waveguide core, their interaction with the surrounding environment is limited, and the Brillouin frequency shift (BFS) is determined by the local conditions within the waveguide \cite{Kurashima1990}. SBS mediated by transverse acoustic waves provides a potential route to partially overcome these limitations, as these waves extend across the entire cross-section of the waveguide \cite{Li2026,Yang2023,layosh2025}. However, they are typically weak and uncommon in conventional optical waveguides, primarily due to poor acoustic confinement and limited optoacoustic overlap, necessitating higher pump power for effective operation. Furthermore, to fully exploit the potential of transverse acoustic wave-mediated Brillouin temperature sensors, the protective jacket must often be removed \cite{wu2025}, thereby increasing their fragility and mechanical vulnerability. SAWs-mediated SBS can be a promising candidate for precise temperature sensing by overcoming these limitations. An artistic representation of SAWs-mediated Stokes scattering is shown in Fig. \ref{fig:Art}(a). SAWs are a combination of longitudinal and transverse acoustic waves that propagate along the surface of the waveguide \cite{ye2025,rodrigues2025,govert2024,simon2025,Lisa2025,klaver2026,Yu2025}. Exploiting SAWs for temperature sensing technologies enables continuous and direct interaction with the surrounding environment, thereby improving measurement accuracy \cite{Kent2024,Ren2025}. Recent studies on the temperature dependence of the BFS in tapered silica fibers have shown that, in addition to the advantages offered by SAWs, the high Brillouin gain achievable in these waveguides significantly reduces the optical power required for efficient SBS operation \cite{maxime2024,Liu_2022,simon2025}. In these structures, the Brillouin frequency exhibits a tunable temperature dependence by adjusting the taper diameter and length. Although these structures provide an exceptionally high Brillouin gain \cite{maxime2024,Liu_2022}, concerns regarding the mechanical stability of such tapered silica fibers might arise.

\par
Here, we present the first comprehensive study of the temperature dependence of intermodal backward SBS in x-cut TFLN-on-insulator by employing both numerical and experimental methods. This study lays the foundation for power-efficient, direct Brillouin temperature sensing technologies based on SAWs. Due to its intrinsic crystal anisotropy, the optical properties of lithium niobate vary with the direction of light propagation, which in turn influences the characteristics of the acoustic waves participating in the SBS process. Taking into account the material anisotropy of lithium niobate, the temperature coefficient of BFS is studied for four different propagation directions, 0$^\circ$, 10$^\circ$, 20$^\circ$, and 30$^\circ$, with respect to the crystallographic y-axis. For the 0$^\circ$-waveguide we measured a temperature coefficient of -0.73 $\pm$ 0.02\,MHz/$^\circ$C and large Brillouin gain of 24.8 $\pm$ 1.3\,m$^{-1}$W$^{-1}$.

\section{Results}

\begin{figure*}[htbp]
\centering
\centerline{\includegraphics[width=1.6\columnwidth, height=0.6\columnwidth]{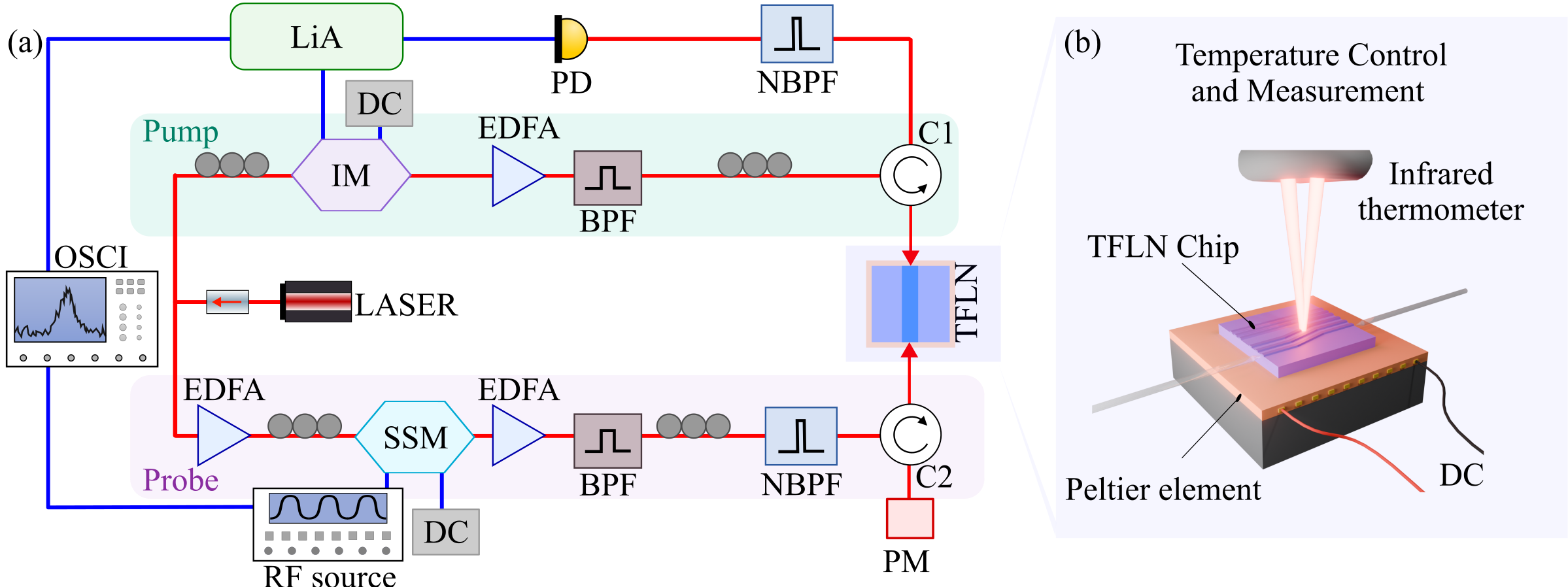}}
\caption{Schematic representation of experimental setup; IM - Intensity Modulator, SSM - Single Sideband Modulator, LiA - Lock-in Amplifier, EDFA - Erbium Doped Fiber Amplifier, BPF - Bandpass Filter, NBPF - Narrow Bandpass Filter, PC - Polarization Controller, PD - Photodetector, OSCI - Oscilloscope, C$_1$ and C$_2$ - Circulators; (b) Technique used to tune temperature and measure surface temperature of the chip. The chip is mounted on a thermoelectric peltier element using double-sided thermal tape, and its temperature is controlled by adjusting the voltage applied to the Peltier element. The surface temperature is monitored using an infrared thermometer.}
\label{fig:1}
\end{figure*}

\begin{figure*}[htbp]
\centering
\centerline{\includegraphics[width=2\columnwidth, height=1.0\columnwidth]{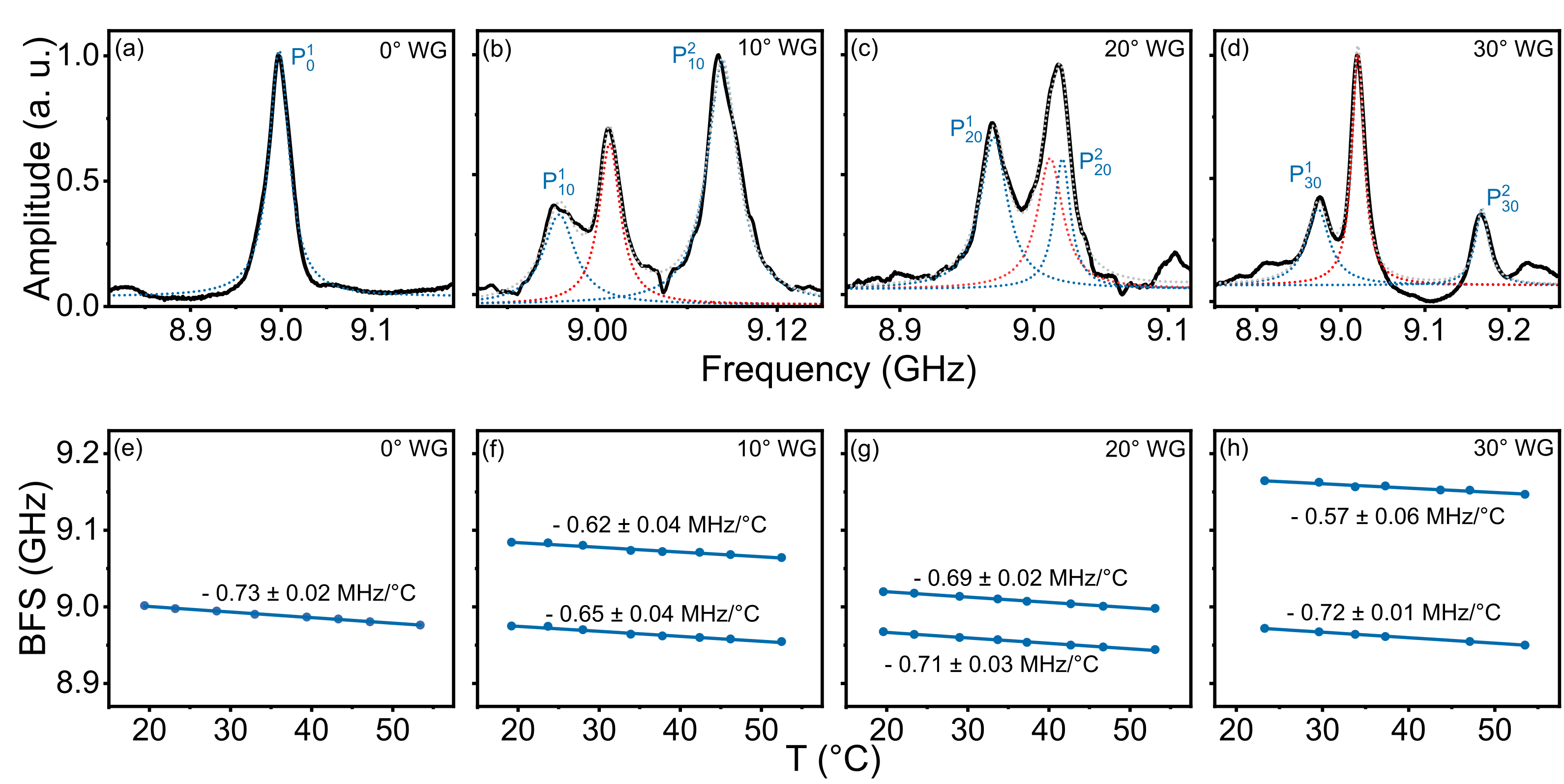}}
\caption{ (a) - (d) Experimental Brillouin spectra at room temperature of 0$^\circ$, 10$^\circ$, 20$^\circ$, and 30$^\circ$-waveguides with Lorentzian fits. The solid and the dotted lines represent experimental data and the Lorentzian fits, respectively; (e) -(h) Measured temperature dependence of BFS for different peaks in 0$^\circ$, 10$^\circ$, 20$^\circ$, and 30$^\circ$ -waveguides. The slope, given for each fit, accounts for the temperature coefficient of BFS.}
\label{fig:2}
\end{figure*}

\begin{figure*}[htbp]
\centering
\centerline{\includegraphics[width=2.05\columnwidth, height=0.48\columnwidth]{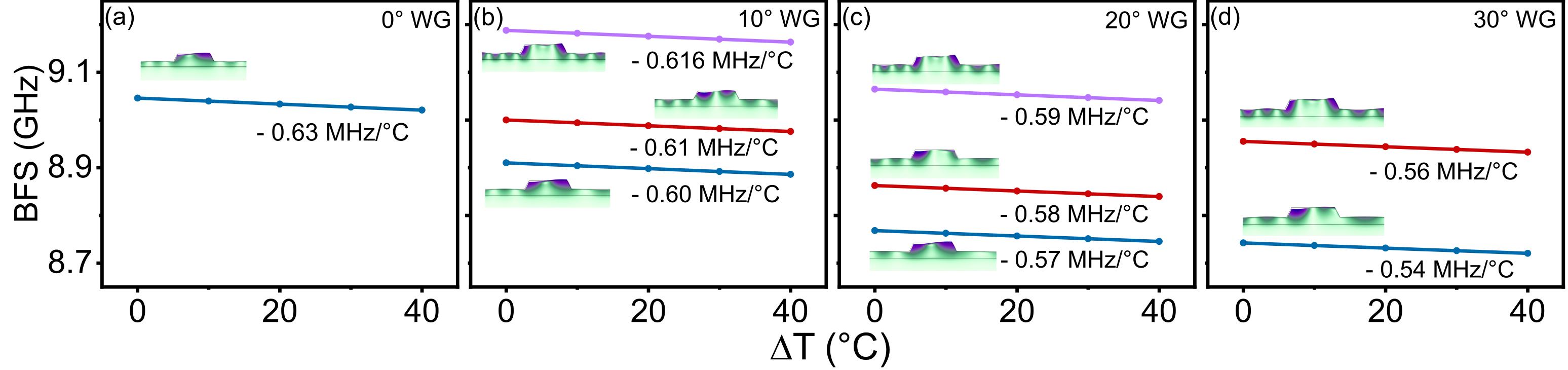}}
\caption{Simulation results for temperature dependence of BFS for (a) 0$^\circ$, (b) 10$^\circ$, (c) 20$^\circ$, and (d) 30$^\circ$ -waveguides. Each graph in the plots is labeled with the corresponding acoustic modes at room temperature.}
\label{fig:1a}
\end{figure*}

\par The waveguides used to perform the experiments are fabricated from a 600\,nm thick x-cut lithium niobate film fabricated by smart cut technique on a 2\,$\upmu$m thick SiO$_2$ layer on top of a 500\,$\upmu$m thick silicon wafer. The lithium niobate is partially etched to a depth of 240\,nm with an etched angle of 28$^\circ$ forming ridge waveguides. An SEM image of the waveguide cross-section is presented in Fig. \ref{fig:Art}(c). Figures \ref{fig:Art}(d) and (e) show the simulated TE and hybrid modes supported by the waveguides. The experiment is performed using a pump-probe setup with a Lock-in amplifier (LiA, HF2LI from Zurich Instruments), as depicted in Fig. \ref{fig:1}(a). The light from a 1550\,nm laser is split into two arms, using a 30/70 coupler to generate pump and probe signals. The upper pump arm, shown in the green rectangle, is modulated with a 20 kHz sine wave from the LiA using an electro-optic intyensity modulator (IM). The IM output is amplified by an Erbium doped fiber amplifier (EDFA), and then the amplified spontaneous emission (ASE) noise from the EDFA is filtered out using a bandpass filter (BPF) to generate the final pump signal. In the probe arm, shown in the purple rectangle, a carrier suppressed single sideband is generated at a frequency, $\omega_L - \Omega_B $, using an single sideband modulator (SSM), where $\omega_L$ and $\Omega_B$ are the laser frequency and Brillouin resonance frequency, respectively. During the measurements, the sideband frequency is swept between $\omega_L - (\Omega_B \pm \Delta\Omega)$. The output of the SSM is amplified using an EDFA, while further ASE noise of the EDFA is filtered using an BPF, to generate the seed signal. A narrow bandpass filter (NBPF) is used in the probe arm to further suppress the other unwanted frequency components. Two lensed fibers are employed to couple light into and out of the waveguide, with a coupling loss of $\sim$5 dB per facet. The Stokes signals are converted into electrical signals with a photodetector (PD) after eliminating the pump reflections using an NBPF. The PD output is connected to the LiA, and the Brillouin responses are measured with an oscilloscope. During each measurement, the pump and probe polarizations are adjusted to maximize the Brillouin gain using the polarization controllers in the pump and probe arms. The photonic chip temperature is tuned between 18 and 55\,$^\circ$C, and the individual Brillouin responses are recorded. Figure \ref{fig:1}(b) illustrates the temperature control and measurement scheme for the chip. To investigate the temperature dependence of SBS for different light-propagation directions in the TFLN, we use waveguides that are fabricated at angles of 0$^\circ$, 20$^\circ$, 10$^\circ$, and 30$^\circ$ with respect to the crystallographic y direction of a single x-cut TFLN chip.

\par Figures \ref{fig:2}(a) - (d) show the normalized Brillouin spectra for different angled waveguides at room temperature. The temperature dependence of the peak positions is depicted in Figs. \ref{fig:2}(e) - (h), where the temperature coefficient of the BFSs can be extracted from the slope of the curves. By comparing with numerical simulations using COMSOL Multiphysics\textregistered, we conclude that these peaks arise from intermodal Brillouin scattering between the TE and hybrid modes. The 10$^\circ$, 20$^\circ$, and 30$^\circ$ -waveguides incorporate 0$^\circ$ -waveguide sections at both ends, facilitating efficient light coupling at the facets (seeFig. \ref{fig:Art}(b)). Consequently, the measured Brillouin spectra for these waveguides are characterized with the contribution from the 0$^\circ$ -waveguide sections, which are highlighted using red colored Lorentzian fits in Figs. \ref{fig:2}(b) - (d). In Figs. \ref{fig:2}(a) - (d), solid black, dotted blue, and dotted grey lines represent experimental data, Lorentzian fits for the individual peaks, and the cumulative fit, respectively. The BFS is measured as a function of the surface temperature of the chip and then linearly fitted (see Figs. \ref{fig:2}(e) - (h)). The slope of the fitted line provides the temperature coefficient of the BFS. In the 0$^\circ$ -waveguide, a prominent peak, $P^1_0$, is observed at frequency 9\,GHz shown in Fig. \ref{fig:2}(a). The measured temperature coefficient of BFS for $P^1_0$ is $-0.73$ $\pm$ 0.02\,MHz/$^\circ$C from Fig. \ref{fig:2}(e). The Brillouin gain for the $P^1_0$ is measured to be 24.8$ \pm $1.3\,m$^{-1}$ W$^{-1}$, determined by comparing it to the Brillouin response of single-mode fiber \cite{LiAGain}. The 10$^\circ$ -waveguide frequency spectra are characterized by two peaks, $P^1_{10}$ and $P^2_{10}$, mediated by two different SAWs with frequencies 8.98\,GHz and 9.08\,GHz as shown in Fig. \ref{fig:2}(b). The temperature coefficient of BFS for $P^1_{10}$ and $P^2_{10}$ are calculated as $- 0.65$ $\pm$ 0.04\,MHz/$^\circ$C and $- 0.62$ $\pm$ 0.04\,MHz/$^\circ$C, respectively, as presented in Fig. \ref{fig:2}(f). Figure \ref{fig:2}(c) shows the Brillouin spectra for the 20$^\circ$ -waveguide. At room temperature, the measured Brillouin spectra consist of two distinct Brillouin peaks denoted as $P^1_{20}$ and $P^2_{20}$, centered at the frequencies 8.97\,GHz and 9.03\,GHz, respectively. The temperature coefficient of BFS for $P^1_{20}$ and $P^2_{20}$, derived from Fig. \ref{fig:2}(g), are $- 0.71$ $\pm$ 0.03\,MHz/$^\circ$C and $- 0.69$ $\pm$ 0.02\,MHz/$^\circ$C, respectively. Similarly, the Brillouin spectra of the 30$^\circ$ -waveguides, shown in Fig. \ref{fig:2}(d) consists of two peaks, $P^1_{30}$ and $P^2_{30}$ located at 8.97\,GHz and 9.17\,GHz, respectively. The corresponding temperature coefficient of BFS, obtained from Fig. \ref{fig:2}(h), are $- 0.72$ $\pm$ 0.01\,MHz/$^\circ$C and $- 0.57$ $\pm$ 0.06\,MHz/$^\circ$C, respectively. The BFS exhibits a negative dependence on temperature, and the maximum temperature coefficient of BFS shift is observed for 0$^\circ$ -waveguides.


\par

For our simulations, we adapted the numerical simulation model originally published by CC Rodrigues \textit{et. al.} \cite{Rodrigues2023,rodrigues2025,wiederhecker2025}, with modifications implemented to consider the effect of temperature and similar geometries as waveguides used for our experiment. In the model, for a given waveguide orientation, the acoustic frequency is determined by the wave vector of the acoustic field. This wave vector is governed by the phase-matching condition essential for SBS, as well as the material properties such as the elastic stiffness tensor and density, which influence the speed of sound in the waveguide. The effect of temperature on the refractive index is taken into account by introducing the thermo-optic coefficients along the ordinary ($dn_{o}/dT$) and extraordinary ($dn_{e}/dT$) axis \cite{Luigi2005}. The temperature dependence of the elastic stiffness tensor \cite{Sevan2022} and density variation (thermal expansion) are incorporated through direct tensor updates. Multiple Brillouin peaks were observed during the simulations; however, weaker interactions were neglected for simplicity. 
\par
Figure \ref{fig:1a} shows the results of simulations for the variation of BFS as a function of temperature change. In Figs. \ref{fig:1a}(a) - (d), the solid blue, red, and violet lines represent the temperature dependence of the BFS in descending order of Brillouin gain. The acoustic modes corresponding to each peak at room temperature are indicated next to their respective graphs. The $0^\circ$ -waveguide displays a strong Brillouin interaction at an acoustic frequency of 9.05\,GHz at room temperature. The variation of the BFS with temperature deviation from room temperature is shown in Fig. \ref{fig:1a}(a). The slope of the plot gives the temperature coefficient of BFS, which is $- 0.63$\,MHz/$^\circ$C. The 10$^\circ$ and 20$^\circ$ -waveguides exhibit a prominent peak accompanied by two secondary peaks at higher frequencies, where the secondary peaks exhibit very poor efficiency. At room temperature, the peaks in the 10$^\circ$ are located at 8.91\,GHz, 9.00\,GHz, and 9.19\,GHz, with corresponding temperature coefficients of $- 0.60$\,MHz/$^\circ$C, $- 0.61$\,MHz/$^\circ$C, and $- 0.62$\,MHz/$^\circ$C, respectively, depicted in Fig. \ref{fig:1a}(b). For the 20$^\circ$ -waveguide, the peaks appear at 8.68\,GHz, 8.86\,GHz, and 9.07\,GHz, with temperature coefficients of $- 0.57$\,MHz/$^\circ$C, $- 0.58$\,MHz/$^\circ$C, and $- 0.59$\,MHz/$^\circ$C, respectively, illustrated in Fig. \ref{fig:1a}(c). The 30$^\circ$ -waveguide consists of a strong interaction at 8.74\,GHz and a weaker interaction at 8.96\,GHz. The corresponding temperature coefficients of the BFS are $- 0.54$\,MHz/$^\circ$C and $- 0.56$\,MHz/$^\circ$C, respectively, as depicted in Fig. \ref{fig:1a}(d). The simulation results indicate that intermodal SBS driven by TE-hybrid interactions is the dominant mechanism across all four angled waveguides, with the 0$^\circ$ -waveguide exhibiting the largest temperature coefficient of SBS. The acoustic modes extracted from the simulations confirm that all Brillouin interactions are mediated by SAWs. For the 10$^\circ$, 20$^\circ$, and 30$^\circ$ -waveguides, multiple Brillouin peaks are observed; however, the secondary peaks at the higher frequencies exhibit very low efficiency, which can be attributed to poor acoustic confinement within the waveguide structure. Previous investigations of Brillouin interactions in TFLN waveguides have shown that these interactions are highly sensitive to the waveguide geometry \cite{Rodrigues2023}. The minor discrepancies between the experimental data and numerical results are attributed to slight deviations in the fabricated waveguide geometry from the ideal design, as well as to inherent real-world imperfections such as surface roughness, fabrication tolerances, and structural inhomogeneities.

\smallskip

\begin{table*}[htbp]
\centering
\caption{\bf Comparison with previous works}
\begin{tabular}{cccc}
\hline
Platform & Type of the acoustic wave &Temperature coefficient (MHz/$^\circ$C) & Gain (m$^{-1}$ W$^{-1}$) \\
\hline

Standard SMF \cite{Xing2018, Kobyakov2010}  & Longitudinal & $+1.19$ & $0.14$ \\
$As_2S_3$ on-chip \cite{Lai2025} & Longitudinal&$-0.55$ & $426$ \\
LiCoF($CF_2$) \cite{geilen2023} &Longitudinal& $-7.5$ & $32.2\pm 0.8$ \\
HNLF \cite{wu2025,Wang2011} & Transverse & $+0.073$ & $0.01735$\\
Tapered silica fiber \cite{simon2025,maxime2024} &Surface& $+2.348$$^a$ &$15$$^b$  \\
\textbf{x-cut TFLN(This work, 0$^\circ$ WG)}&\textbf{Surface} & \textbf{$- 0.73$ $\pm$ $0.02$} & \textbf{$24.8$$ \pm $$1.3$} \\
\hline
\end{tabular}
  \label{tab:1}
  
$^\textit{a}$Taper diameter = 710 nm, taper length = 100 mm; $^\textit{b}$Taper diameter = 740 nm, taper length = 100 mm
\end{table*}

\section{Discussions and Conclusions}
In this work, we experimentally and numerically investigated the temperature coefficient of the BFS for different light propagation directions in x-cut TFLN waveguides. Our results demonstrate that the waveguides support strong intermodal SBS between TE and hybrid modes, mediated by SAWs. The proposed x-cut TFLN platform combines direct interaction of SAWs and the surrounding environment and a Brillouin gain of 24.8 $\pm$ 1.3\,m$^{-1}$W$^{-1}$, while exhibiting a BFS temperature coefficient of $-0.73 \pm 0.02$\,MHz/$^\circ$C for the 0$^\circ$ -waveguide. Table \ref{tab:1} compares the performance of the proposed platform with existing state-of-the-art Brillouin temperature sensing platforms. Existing Brillouin temperature sensing schemes, including on-chip $As_2S_3$\cite{Lai2025} and liquid-core fibers (LicoF)\cite{geilen2023}, in which longitudinal acoustic waves are involved, achieve high Brillouin gains of 426\,m$^{-1}$W$^{-1}$ and $32.2 \pm 0.8$\,m$^{-1}$W$^{-1}$, respectively. However, these platforms lack direct contact between the acoustic waves and the environment. In contrast, platforms such as highly nonlinear fiber (HNLF)\cite{Wang2011}, which involve transverse acoustic waves, provide slightly greater exposure of the acoustic waves to the environment but suffer from a very low Brillouin gain of 0.01735\,m$^{-1}$W$^{-1}$. SAW-mediated SBS in tapered silica fibers \cite{maxime2024} provides a Brillouin gain of 15\,m$^{-1}$W$^{-1}$ while offering direct contact to the surrounding temperature. Nevertheless, their mechanical durability remains a significant limitation. Compared with existing platforms, the proposed technique overcomes these limitations by enabling direct exposure of the acoustic waves to the environment while maintaining a high Brillouin gain, rendering it a versatile platform for temperature sensing applications. This approach provides an optically power-efficient and in-situ temperature-sensing solution for integrated Brillouin temperature sensors with broad applicability in areas such as the medical field \cite{Ferenc2009,Li2022}, process-controlled manufacturing \cite{woo2009}, and automobiles \cite{kerttula2024}.

\section{Acknowledgment}
  The authors would like to thank Abdullah Alabbadi from the Microphotonics Research Group, MPL, for his assistance in capturing the SEM image of the sample and Ernst-Lukas Kuhlmann from the Integrated Quantum Optics Group, Paderborn University for his help in designing the waveguide mask. We acknowledge funding from the PhoenixD Cluster of excellence EXC 2122 by the German Research Foundation, Max Planck Society through the Independent Max Planck Research Groups scheme and the MPC Kick starter grant. P.K. thanks the support from the  International max Planck research school (IMPRS PL). G.J-S. acknowledges sponsorship by the Alexander von Humboldt Foundation via the Humboldt Research Fellowship Program for Postdocs. S. B. acknowledges the Max Planck School of Photonics supported by the Dieter Schwarz Foundation, the German Federal Ministry of Research, Technology and Space (BMFTR), and the Max Planck Society. 
\nocite{*}
\bibliography{ref}

\end{document}